\documentclass[9pt,twocolumn,twoside]{opticajnl}
\journal{opticajournal} % use for journal or Optica Open submissions

\setboolean{shortarticle}{true}
\usepackage{lineno}
\usepackage{subfigure}
\title{%Sub-2-cycle multi-milliJoule single-stage post-compression in a multipass cell
TW-class sub-2-cycle post-compression of Ti:Sapphire laser pulses in a gas-filled multipass cell
}

\author[1,*]{Louis Daniault}
\author[1]{Jaismeen Kaur}
\author[2]{Geoffrey Gallé}
\author[2]{Cedric Sire}
\author[2]{François Sylla}
\author[1]{Rodrigo Lopez-Martens}

\affil[1]{Laboratoire d'Optique Appliquée (LOA), Institut Polytechnique de Paris, ENSTA Paris - CNRS - Ecole Polytechnique, 91120 Palaiseau, France}
\affil[2]{SourceLAB, 7 rue de la Croix Martre, 91120 Palaiseau, France}

\affil[*]{louis.daniault@ensta-paris.fr}

\begin{abstract}
We report on the nonlinear temporal post-compression of 7~mJ sub-40~fs pulses from a commercial kHz Ti:Sapphire laser down to a record 4~fs duration (1.5 optical cycle) in a compact single-stage gas-filled multi-pass cell, with 60\% overall compression efficiency.

\end{abstract}

\setboolean{displaycopyright}{false} % Do not include copyright or licensing information in submission.

\begin{document}

\maketitle

%\section{Introduction}
In recent years, multi-pass cells (MPC) have been increasingly used for post-compressing ultrashort laser pulses and boosting their peak power for applications~\cite{Nagy:20,Viotti:22}. Indeed, MPCs exhibit high compression ratios, high energy throughput, low footprint, reduced complexity, and remarkable spatio-temporal pulse quality \cite{Schulte:16,Hanna:17,Pfaff:23,Grebing:20,Escoto:22}. The vast majority of MPC-based experiments reported so far have involved Ytterbium (Yb)-based laser systems, for which the pulse duration was reduced from the picosecond (ps) level down to a few tens of femtoseconds (fs), thereby boosting their peak power by more than one order of magnitude. However, key applications, such as attosecond pulse generation from plasma mirrors~\cite{Jahn:19,ouille_lightwave-controlled_2024} or laser-plasma electron acceleration~\cite{Faure:19}, require high-energy few-cycle pulses that are much more challenging to produce. Indeed, their manipulation requires the use of low-dispersion broadband metallic mirrors or dispersion-matched optics with lower damage threshold, which increases both the footprint and complexity of the MPC design. Generating few-cycle pulses from Yb-based lasers typically involves at least two distinct post-compression stages: the first, compact, efficient, and adapted to moderate spectral bandwidths, and the second, bulkier, more lossy and input sensitive, bearing most of the system complexity. Recently, multi-mJ few-cycle pulses were post-compressed in such a two-stage gas-filled MPC configuration down to 9.6~fs, corresponding to roughly 3 optical cycles and a fraction of TW peak power~\cite{Rajhans:23}.

MPC-based post-compression has also been demonstrated with amplified short-pulse Ti:Sapphire lasers, producing few-10~fs pulses that can be directly post-compressed down to few-cycle duration in a single stage. This was first demonstrated at the 100\,$\mu$J-level~\cite{Rueda:21} down to 3 optical cycles and then at the mJ-level~\cite{Daniault:21} down to 2 cycles. In all these experiments, the compression factor was limited by the dispersion induced by the nonlinear medium itself and the minimal achievable pulse duration to approximately 2 cycles. In this work, we report on the post-compression of 7~mJ energy sub-40~fs pulses from a 1~kHz Ti:Sapphire laser down to 4~fs, corresponding to 1.5~optical cycle at 800~nm central wavelength, in a 3~m long single-stage helium-filled MPC with 60\% overall compression efficiency, performance on par with that of hollow-fiber compressors~\cite{Nagy:20}.

%\section{MPC design}
%The large amount of chromatic dispersion accumulated in a gas-filled MPC (tens of meters of propagation length) progressively reduces the pulse peak power available for nonlinear spectral broadening and therefore limits the minimal achievable output pulse duration. Dispersion-engineered MPCs involving spherical chirped end-mirrors can be used to combat this saturation effect~\cite{Rajhans:23} but the total dispersion has to be tailored to the specific MPC design, leaving little flexibility and restricting MPC operation to a fixed point. One may also introduce controlled dispersion in the MPC to smooth the broadened pulse spectrum in an attempt to privilege temporal pulse quality over pulse duration~\cite{Omar:24}. 

In this experiment, the input pulses carry seven times more energy than in our previous work~\cite{Daniault:21}. Using a home-made numerical MPC designer to simulate nonlinear pulse propagation in space and time, we managed to keep the same MPC footprint of around 3~m in length and find a stable Eigenmode with 4.3\,mm beam diameter on the 1.5\,m ROC end-mirrors, corresponding to a peak fluence below 100 mJ/cm$^2$, which is still within the safe zone specified for the enhanced silver mirror coatings. This geometry allows a maximum of 38 passes for 101.6\,mm diameter cavity end-mirrors. Such a highly concentric cell, however, yields a tighter beam waist at the center of the MPC, estimated at 350~$\mu$m $1/e^2$ diameter, which can lead to drastic ionization effects. Under these conditions, simulations indicate that the input pulses from the Ti:Sapphire laser can be post-compressed down to about 4~fs pulses using helium gas at a pressure of 2 bar and circular laser polarization, with negligible ionization effects. These simulation results provided the guidelines for designing the experiment.

%\section{Experimental setup}

\begin{figure}[ht]
\centering
\includegraphics[width=.9\linewidth]{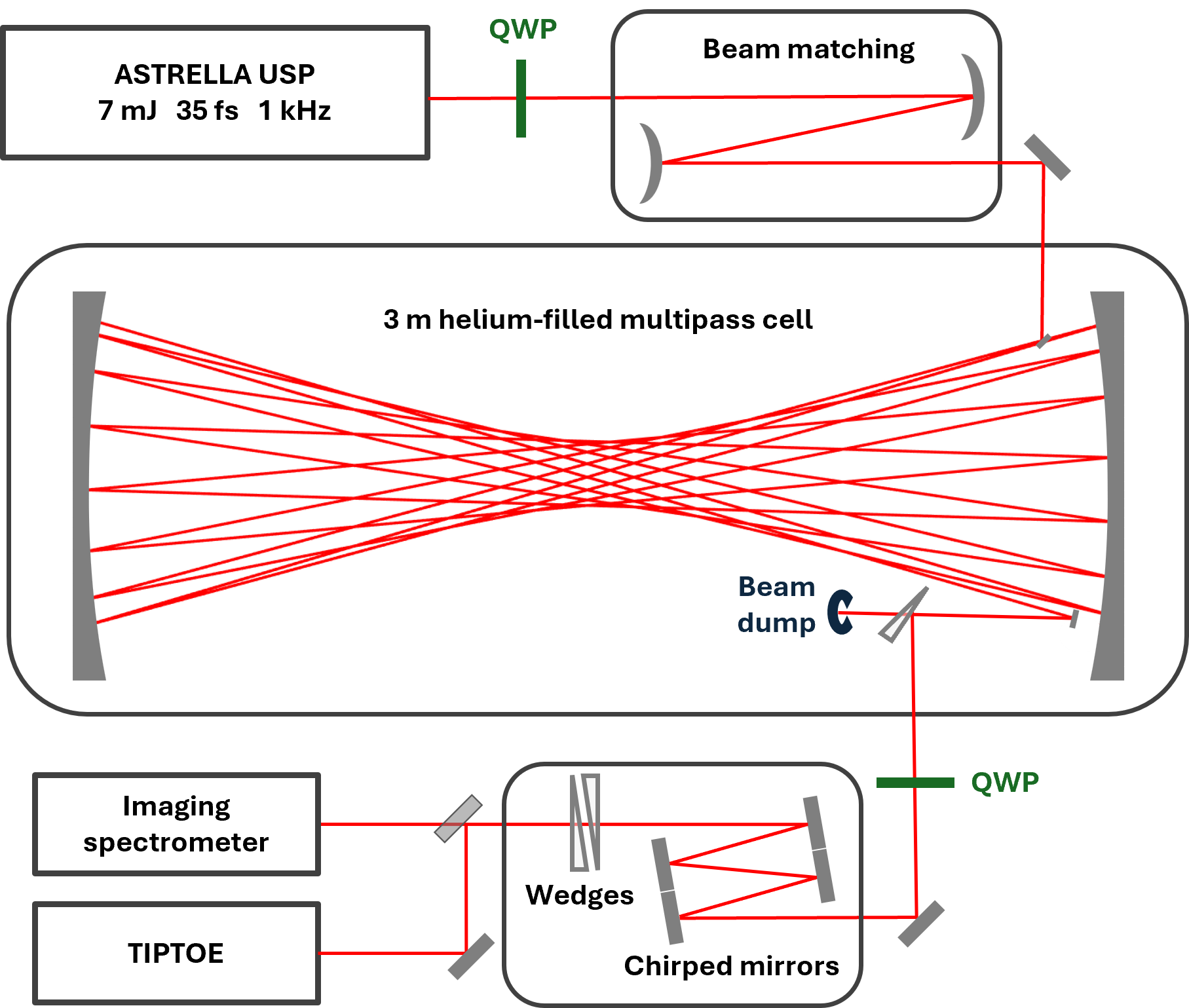}
\caption{MPC post-compression setup for the ASTRELLA USP Ti:Sapphire laser.}
\label{setup}
\end{figure}

The experimental setup is depicted in fig.\ref{setup}. The Ti:Sapphire laser is an ASTRELLA USP (Coherent) that delivers up to 7~mJ energy 35~fs pulses at 1~kHz repetition rate. The MPC has been set up exactly according to the aforementioned design and placed inside a vacuum chamber. A first quarter-wave plate (QWP) allows the MPC to be operated with circular polarization, which also reduces any nonlinearities experienced in air prior to injection into the vacuum chamber. A pair of spherical mirrors are used to match the laser beam to the Eigenmode of the MPC. Even though the pulse peak power is seven times above the critical power in air for circular polarization, the beam is large enough such that the slight onset of self-focusing can be accommodated and taken into account for mode matching. The situation is drastically different at the output of the MPC. When coupled back out into air, the beam is small enough such that self-focusing prevents it from further diverging, and the beam quickly collapses. Therefore, a wedged window is placed in the vacuum chamber just before the beam exits the vacuum chamber in order to sample a small fraction of the pulse energy and to be able to characterize the compressed pulses in air.

The attenuated output beam is then collimated with a spherical mirror, and a second QWP sets the laser polarization back to linear horizontal. The compressor that follows consists of several pairs of double-angle chirped mirrors with -40 fs$^2$ group delay dispersion per bounce (PC70 from Ultrafast Innovations), and pulse compression is finely tuned using a pair of AR-coated fused silica wedges. The beam is finally characterized temporally with a TIPTOE~\cite{Cho2021} (SourceLab) and spatially with an imaging spectrometer (MISS from FemtoEasy). 

The MPC throughput is characterized for 28 passes by measuring the output power at full input power and by highly chirping the input pulses, which simulates linear propagation inside the MPC just as if the beam was propagating through the cell under vacuum. For 6.7~mJ effective pulse energy sent into the MPC (measured just before the injection mirror), we obtained 4.6~mJ energy at the output, measured just before the wedged window, leading to an MPC throughput of 69$\%$, consistent with the end-mirror reflectivity and the number of passes ($\sim 99\%$ per bounce). When the MPC is set under vacuum, the output power of the sampled beam measured outside the cell remains unchanged compared to when the MPC is filled with gas, showing that the nonlinear spectral broadening process barely affects the MPC throughput.

\begin{figure}[ht]
\centering
\includegraphics[width=.9\linewidth]{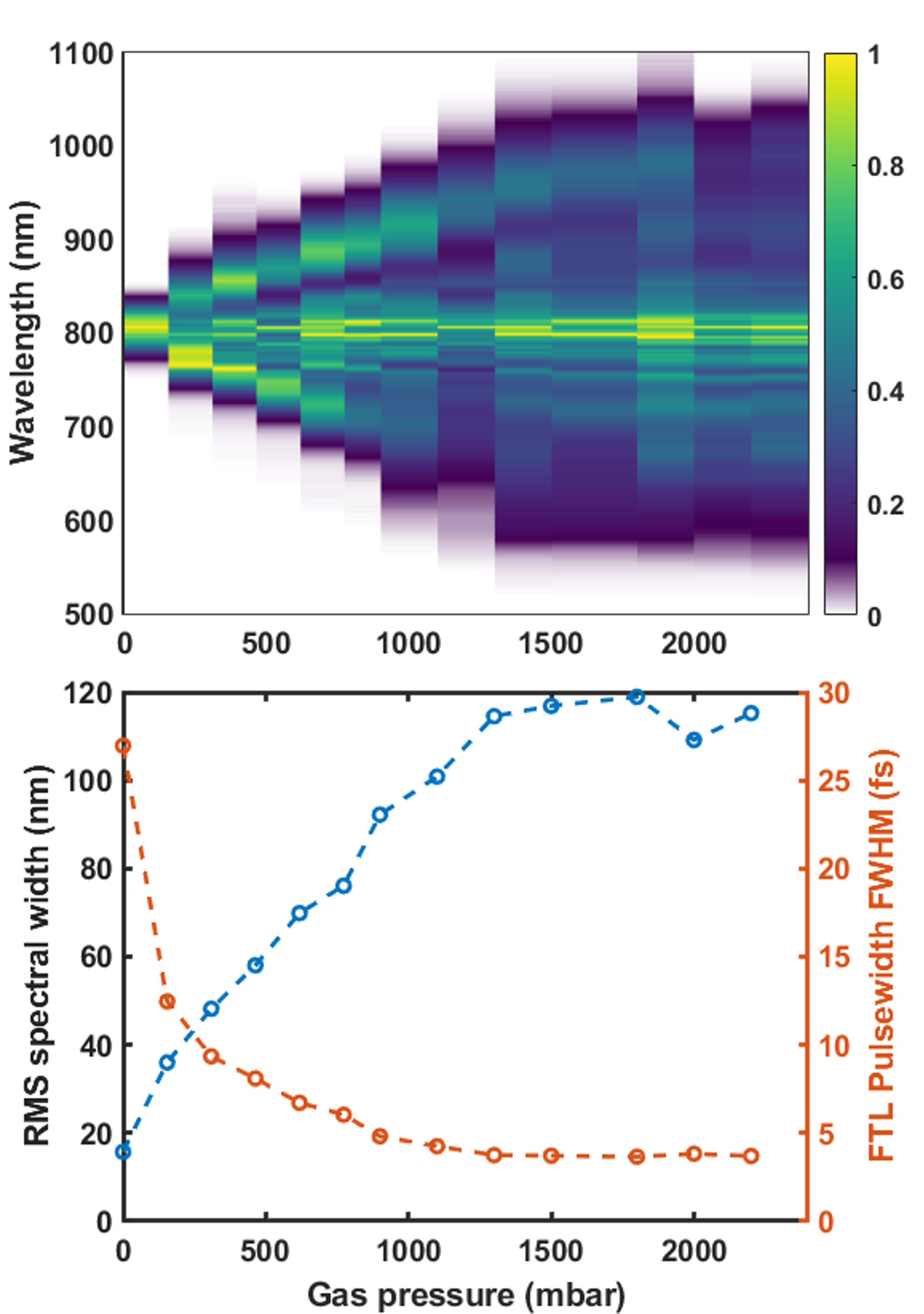}
\caption{Evolution of the measured output spectrum (top) and the corresponding computed RMS spectral width and FTL pulse duration (bottom) as a function of increasing gas pressure inside the MPC.}
\label{SpecEvol}
\end{figure}

The output spectrum was recorded for increasing gas pressures from 0 to 2200~mbar, as shown in fig.~\ref{SpecEvol}, along with the computed RMS spectral width and the corresponding Fourier transform-limited (FTL) duration. It is worth noting that the input FTL pulse duration is actually shorter than 35~fs, due to incompressible residual higher-order spectral phase. At low pressures, the spectral width increases linearly, as expected for dispersion-free spectral broadening, up to approximately 1200~mbar, where the broadening begins to saturate. This marks the onset of the smoothing effect of linear dispersion, as evidenced by the gradual disappearance of the modulations observed in the spectrum above 1200~mbar. At gas pressures exceeding 1800~mbar, the spectrum becomes more distorted and peaked, with a corresponding RMS spectral width exhibiting chaotic behavior. At these pressures, increasing the nonlinearity no longer leads to further spectral broadening but instead significantly distorts the spectral profile and phase. This was confirmed by the poor pulse compressibility and distorted output temporal pulse profiles with durations deviating from the FTL value. In addition, such high gas pressures introduce even more group-delay dispersion to be compensated in the compressor, which unnecessarily increases losses and complexity. These spectral distortions can be considered as the early signs of pulse breakdown caused by excessive self-steepening, which arises at even higher pressures, above 2200~mbar, as reported in our previous work \cite{Daniault:21}. 

Following this comprehensive study, we set the Helium pressure to the optimal value of 1800~mbar. The spectral broadening is then optimized by carefully cancelling both the residual input second- and third-order dispersion thanks to the ASTRELLA USP compressor tuning capabilities. The output pulses were fully recompressed using 8 pairs of chirped mirrors, leading to an overall dispersion of -640 fs$^2$ and 87$\%$ transmission. The overall post-compression efficiency is then 60$\%$, and the MPC could potentially deliver more than 4~mJ if the full-energy beam was compressed under vacuum.

\begin{figure}[ht]
\centering
\includegraphics[width=\linewidth]{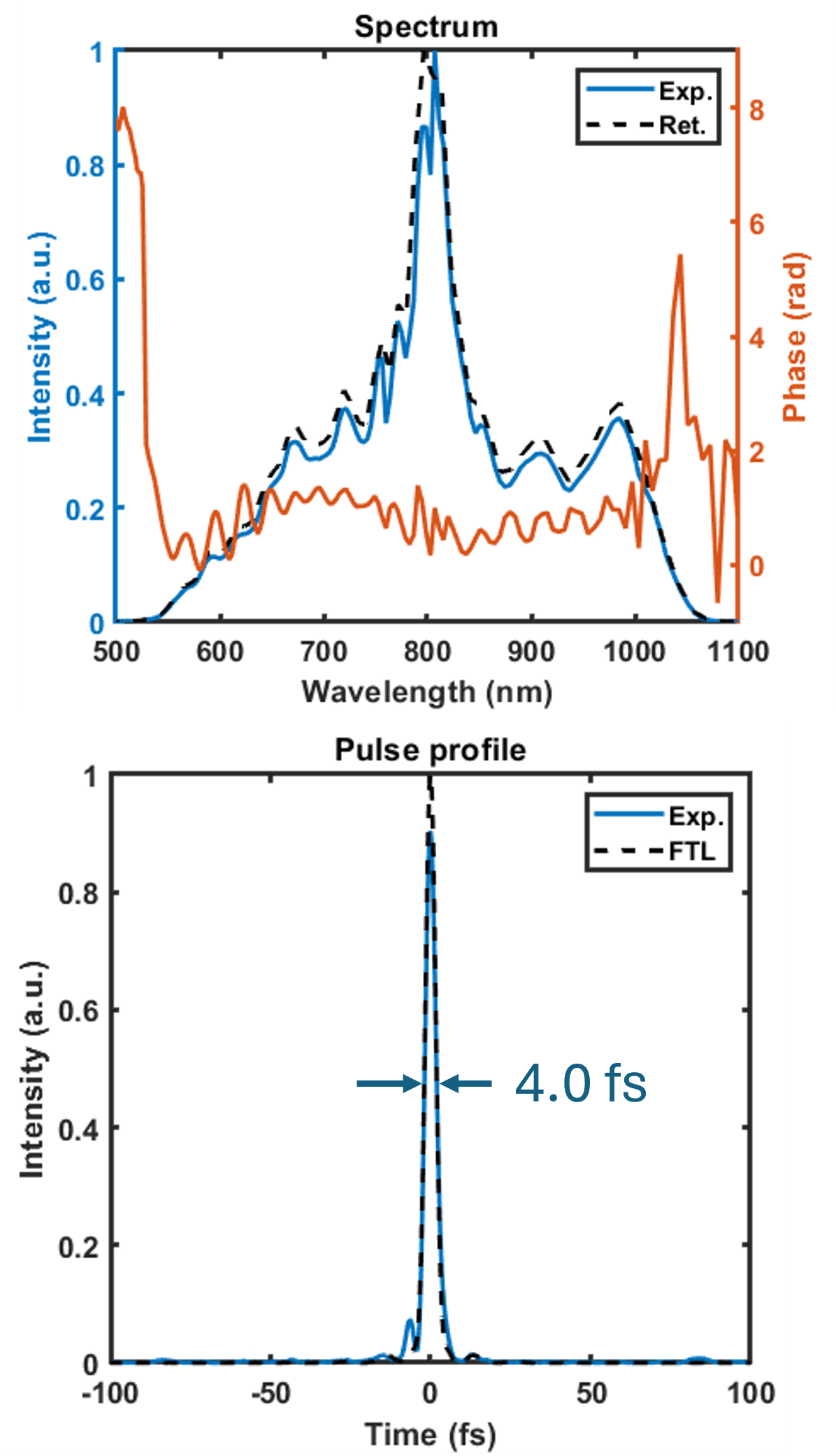}
\caption{TIPTOE measurements. Top: reconstructed spectral intensity (left) and phase (right), along with pulse spectrum measured with a spectrometer (left). Bottom: retrieved pulse temporal profile and computed FTL shape.}
\label{TIPTOE}
\end{figure}

Fig.~\ref{TIPTOE} shows the spectral and temporal properties of the compressed pulses measured with the TIPTOE. The output duration is measured to be 4.0~fs, that is 1.5 cycle at 800~nm central wavelength, with a temporal Strehl ratio of 90$\%$ compared to its 3.7~fs FTL derived from the retrieved spectrum. This is, to the best of our knowledge, the shortest pulse ever generated from an MPC and compression under vacuum at full energy would provide TW-peak-power pulses at 1~kHz repetition rate at the near-single-cycle level. 

The TIPTOE retrieval algorithm can be run using a pulse spectrum measured by an independent spectrometer, which explains the excellent agreement between measured and retrieved spectra. However, letting the retrieval algorithm run freely leads to a very similar spectral shape and identical pulse temporal profile, with comparable retrieval errors~\cite{Cho2021}. The results are shown here using the forced retrieval, which appears to be more accurate as it is better correlated to the external measurement. The retrieval accuracy is further confirmed by the clearly visible residual oscillations in the spectral phase, which nicely correspond to those induced by the double-angle chirped mirror pair arrangement. The main central peak at 800~nm observed in the broadened spectrum can be attributed to the incompressible residual high-order spectral phase of the input pulse.

\begin{figure}[ht]
\centering
\includegraphics[width=\linewidth]{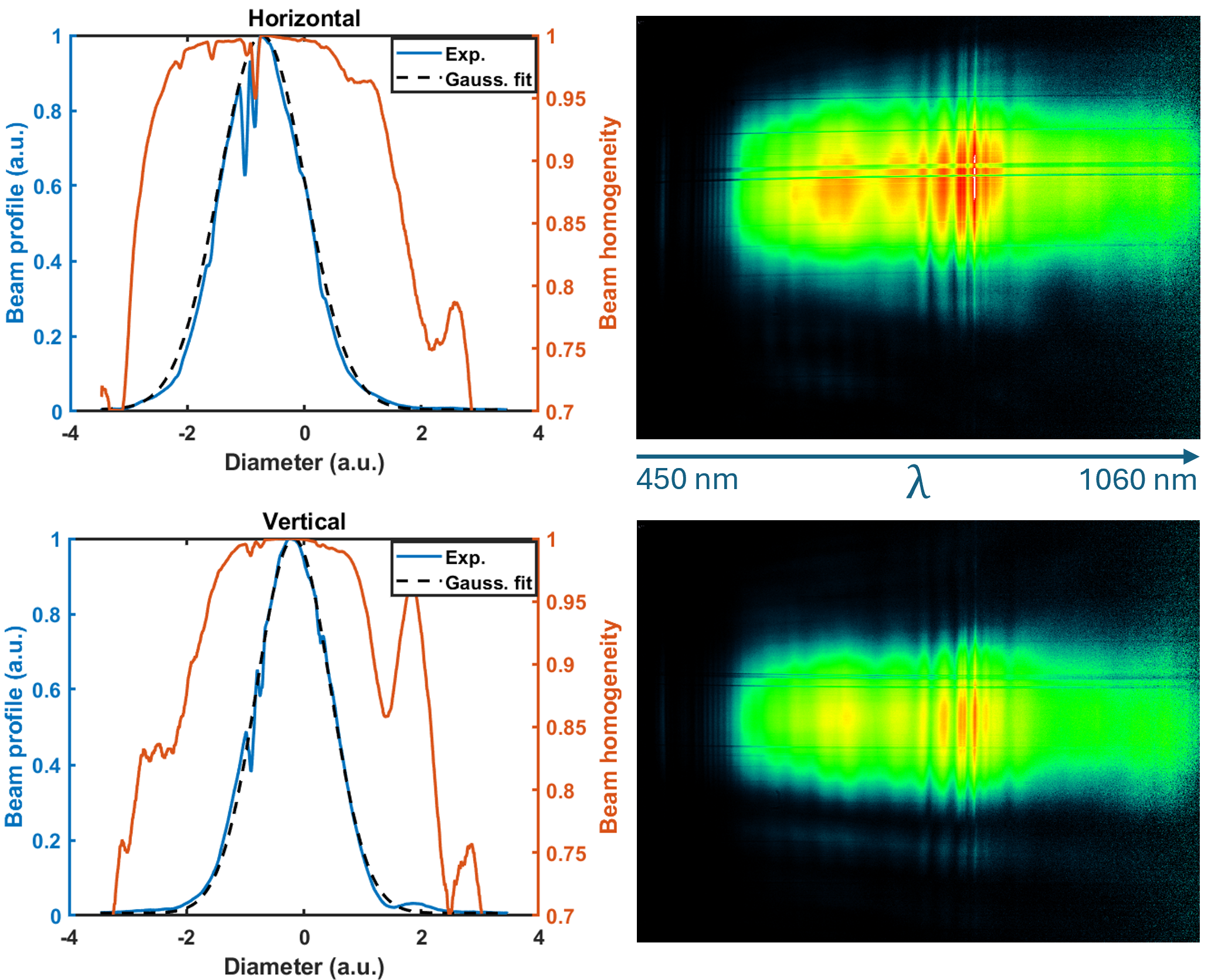}
\caption{Spatio-spectral intensity distribution of the pulses measured with an imaging spectrometer along horizontal (top) and vertical (bottom) beam axes. Left: overall beam profile with spatial homogeneity coefficient (V-parameter, red). Right: raw spatio-spectral measurement.}
\label{MIIS}
\end{figure}

Fig.\ref{MIIS} shows the spatio-spectral homogeneity measured in the far-field along both horizontal and vertical beam axes with the MISS imaging spectrometer. The frequency-summed beam profiles exhibit a Gaussian-like shape with spectral homogeneity above 99$\%$ at FWHM and above 97$\%$ at 1/$e^2$ on both axes. These results demonstrate how MPCs can provide excellent spatio-temporal pulse quality even for nearly octave-spanning spectra.

In conclusion, we demonstrate the scaling of gas-filled MPC-based post-compression of Ti:Sapphire laser pulses with up to 7~mJ of energy, achieving record 1.5~cycle post-compressed pulse durations of 4~fs at 1~kHz repetition rate. Compression under vacuum would provide the basis for TW-class laser near-single-cycle light source with reduced complexity and footprint. The output pulse duration and system throughput can be further improved through careful dispersion management inside the cell, which would allow even higher output peak power. MPC-based post-compression of commercial high-average-power laser systems paves the way towards real-life applications of high-flux secondary particle and radiation beams produced from intense laser-plasma interactions.

\begin{backmatter}

\bmsection{Funding} This project has received funding from the European Union’s Horizon 2020 Research and Innovation programme under Grant Agreement No 101004730.

\bmsection{Acknowledgments} The authors thank Jean-Luc Tapié and Coherent Inc. for their support during these experiments.

\bmsection{Disclosures} The authors declare no conflicts of interest.

\bmsection{Data availability} Data underlying the results presented in this paper are not publicly available at this time but may be obtained from the authors upon reasonable request.

\end{backmatter}

% Bibliography
\bibliography{sample}

% Full bibliography added automatically for Optics Letters submissions; the following line will simply be ignored if submitting to other journals.
% Note that this extra page will not count against page length
\bibliographyfullrefs{sample}

%Manual citation list
%\begin{thebibliography}{1}
%\bibitem{Zhang:14}
%Y.~Zhang, S.~Qiao, L.~Sun, Q.~W. Shi, W.~Huang, %L.~Li, and Z.~Yang,
 % \enquote{Photoinduced active terahertz metamaterials with nanostructured
  %vanadium dioxide film deposited by sol-gel method,} Opt. Express \textbf{22},
  %11070--11078 (2014).
%\end{thebibliography}

\end{document}